\documentclass[12pt,preprint]{aastex}
\slugcomment{Not to appear in Nonlearned J., 45.}

\shorttitle{SMBHB evolution and acceleration of jet precession}
\shortauthors{Liu \& Chen}

\begin{document}

\title{Evolution of Supermassive Black Hole Binary and Acceleration of
  Jet Precession in Galactic Nuclei}

\author{F.K. Liu and X. Chen}
\affil{Astronomy Department, Peking University, 100871 Beijing, China}
\email{fkliu@bac.pku.edu.cn, chenx@bac.pku.edu.cn}

\begin{abstract}
    Supermassive black hole binary (SMBHB) is expected with the
    hierarchical galaxy formation model. Currently, physics processes 
    dominating the evolution of a SMBHB are unclear. An interesting
    question is whether we could observationally determine 
    the evolution of SMBHB and give constraints on the physical
    processes. Jet precession have been observed in many AGNs and
    generally attributed to disk precession. In this paper we
    calculate the time variation of jet 
    precession and conclude that jet precession is accelerated in
    SMBHB systems but decelerated in others. The acceleration of jet
    precession $dP_{\rm pr} / dt$ is related to jet precession
    timescale $P_{\rm pr}$ and SMBHB evolution timescale
    $\tau_{\rm a}$, ${dP_{\rm pr} \over dt} \simeq - \Lambda {P_{\rm
    pr} \over \tau_{\rm a}}$. Our calculations based on the models for
    jet precession and SMBHB evolution show that $dP_{\rm pr} / dt$
    can be as high as about $- 1.0$ with a typical value $-0.2$ and
    can be easily 
    detected. We discussed the differential jet precession for NGC1275 
    observed in the literature. If the observed rapid acceleration of
    jet precession is true, the jet precession is due to the orbital
    motion of an unbound SMBHB with mass ratio $q\approx 0.76$. When
    jets precessed from the ancient bubbles to the currently active jets,
    the separation of SMBHB decrease from about $1.46 \, {\rm
    Kpc}$ to $0.80 \, {\rm Kpc}$ with an averaged decreasing velocity 
    $da/dt \simeq - 1.54 \times 10^6 \, {\rm cm/s}$ and evolution 
    timescale $\tau_{\rm a} \approx 7.5\times 10^7 \, {\rm
    yr}$. However, if we assume a steady jet precession for many
    cycles, the observations implies a hard SMBHB with mass ratio
    $q\approx 0.21$ and separation $a\approx 0.29 \, {\rm pc}$.

\end{abstract}


\section[]{Introduction}
\label{sec:intr}

    In the hierarchical galaxy formation models of cold dark matter
    (CDM) cosmology, present-day galaxies are the products of
    successive mergers. Recent observations show that almost all
    galaxies harbor at center 
    a supermassive black hole (SMBH) of mass tightly correlating with 
    both the mass and the velocity dispersions of the bulge 
    \citep{ferrarese00,gebhardt00,magorrian98,tremaine02}. During
    galaxy interacting and merging, the gas at galactic plane is
    driven toward central SMBH, triggering the activity of active
    galaxy \citep{wilson95} and black hole accretion. SMBHs in
    galactic nuclei likely increase mainly through matter accretion.
    In this scenario,
    galaxy interacting and merging is expected to trigger the
    formations of an unbound binary active galactic nuclei (AGNs).
    In galaxy mergers, two galaxies and the SMBHs at center
    initially lose their orbital angular momentum owing to galactic
    dynamic friction and form a bound supermassive black hole binary
    (SMBHB) at a separation of $a_{\rm H} \sim 10 \, {\rm pc}$, when
    the SMBHB bind energy becomes dominated. The dynamics friction is
    very efficient because of the traps of stellar objects around each
    black hole and the evolution timescale of SMBHB is order of the
    local dynamic timescale, depending on the stellar velocity of two
    galaxies. The evolution of a bound SMBHB is dominated by the
    dynamic friction but the evolution timescale depends on the inner
    surface brightness profiles of galaxies. A SMBHB becomes hard
    at a separation $a_{\rm h} \sim 0.1 - 1 \, {\rm pc}$, when the
    loss of the orbital angular momentum is dominated by three-body
    interactions between SMBHB and the stars passing by
    \citep{begelman80,quinlan96,yu02}.  
    When SMBHB becomes hard but orbital angular momentum loss because
    of gravitational wave radiation is unimportant, SMBHB may stall at
    $a\sim a_{\rm h}$ on a timescale longer than the
    Hubble time. However, observations of nearby galaxies suggest that
    most SMBHBs should have passed through the hard phase and
    become coalesced quickly. This is the so-called {\it final parsec
    problem} \citep{merritt05}. To solve the problem, theoretically
    several processes with large uncertainties have been suggested in
    the literature \citep{merritt05} and the hydrodynamics interaction
    with  gas disk may play the major role \citep{gould00,liu03,liu04,
      armitage05,escala04}. An important question is whether we could
    detect SMBHB at galactic center, determine its evolution, and give
    observational constraints on the formation and evolution of
    SMBHB. 

    Although unbound AGN binary systems with separation of order Kpc or
    larger have been imaged in interacting and merging galaxies
    \citep[e.g.][]{komossa03,ballo04,rodr06}, no hard or
    bound SMBHB has been directly detected. Close SMBHBs or binary
    coalescence have been 
    introduced in explaining the observations of many AGNs, for example,
    periodic optical and radio outbursts \citep{sillanpaa88,katz97,
      liu95,liu97,liu06,liu02}, periodic variation
    of Very Long Baseline Interferometer (VLBI) jet position angle
    \citep{stirling03,sudou03}, the interruption of jet formation in 
    double-double radio galaxies (DDRGs) \citep{liu03}, X-shaped radio
    feature in winged radio sources \citep{liu04,merritt02}, and 
    S- or Z-shaped morphological symmetry of radio jets
    \citep[BBR]{begelman80}. A recent review on the observational
    evidences for SMBHB was given by \citet{komossa06}. The periodic
    outbursts may be due to the periodic interaction of a SMBHB and
    a standard accretion disk or an ADAF, while the periodic
    variation of VLBI jet position angle is because of binary orbital
    motion.  However, when the disk mass inside the binary orbit is
    less than the mass of the secondary which is at a radius order of
    $10^3$ 
    times the Schwarzschild radius of the primary black hole, the
    interaction between the accretion disk and the SMBHB will realign
    the inner accretion disk and the binary orbital plane
    \citep{ivanov99,liu04}. When a SMBHB becomes coalecing due to
    gravitational wave radiation, the interaction between the secondary
    black hole and the aligned accretion disk would remove the inner
    disk region and leave a truncated outer accretion disk, leading to
    the interruption of jet formation in DDRGs \citep{liu03} and to
    the formation of delayed X-ray afterglow of a gravitational wave
    radiation burst \citep{milos05}. 

    The S- or Z-shaped radio morphological symmetry has been observed
    for a high fraction of AGNs and was suggested as 
    an observational evidence of SMBHB by \citet{begelman80}. It is
    explained with jet precession of periods about 
    between $10^3$ and $10^8 \, {\rm yr}$ \citep[e.g.][]{gower82,
      hutchings88,dunn06}, because of geodetic precession of the spin
    axis of the primary rotating SMBH misaligned with binary total
    angular momentum \citep{begelman80,roos88}, the orbital motion of
    jet ejecting black hole, the disk precession tidally perturbed by the
    secondary black hole \citep{katz97,dunn06}, or the precession of a
    inner warped disk owing to Bardeen-Peterson
    \citep[e.g.][]{lu05,caproni06}. In all the models reviewed above,
    it has implicitly assumed that the precession of jet orientation 
    follows the precession of the spin axis of the emitting
    rotating black hole and the rotating axis of the inner region of
    accretion disk around the black hole. Although there are many
    models for the jet formations, it is usually believed that jets
    form in the inner disk region along the spin axis of black hole or
    the rotating axis of the inner region of accretion disk, depending
    on the driving energy resources. Because the small characteristic
    size of jet production region \citep[e.g.][]{meier01} and the
    alignment of rotating black hole and the inner region of
    accretion disk due to Bardeen-Peterson effect \citep{bardeen75},
    it seems reasonable to assume that jet would always
    orient along the rotating axis of both the black hole and the
    inner regions of accretion disk irrespective of the driving
    mechanisms. Thus, in this paper we will take the same assumption
    that jets would, if present, precess with the rotating axis of the 
    emitting central black hole and the accretion disk. With the
    assumption, all the models in the literature can explain the
    observed jet precession, though the results 
    depends on many parameters with very large uncertainties. One of
    the very important question is how to tell which model is the
    right one and to determine parameters. With the improvement of
    observational instrument, jet precession can be observed up to
    many cycles, which make it possible to measure the precession
    timescale with very high accuracy. In this paper, we investigate
    the possibility to measure 
    the time derivative of jet precession timescale. In the precession
    models, a rigid-body like disk precession is assumed in the
    literature. Here we take the same assumption.

    A circumbinary accretion disk could be warped by a massive SMBHB
    with random orbital inclination angle relative to accretion disk
    \citep{liu04,ivanov99}. The interaction quickly realigns the inner
    warped disk region and finally the central rotating SMBH with the
    binary orbital plane, while the outer unperturbed disk region far
    from the binary orbit remains coplanar with the 
    galactic plane. This scenario predicates the
    formations of X-shaped radio feature in FRII radio galaxies and a
    random distribution of jet orientation with respect to the
    galactic plane \citep{liu04}. A warped disk precesses, leading to
    jet precession. Therefore, before discussing the variations
    of jet precession timescale, we calculate the precession
    of a warped circumbinary non-massive disk in this paper. Although 
    the detailed SMBHB models for jet precession are different, all of
    them predicates a increase of precession timescale with binary
    evolution. As SMBHBs in galactic nuclei never get softer
    \citep{quinlan96}, the secondary black hole always migrates toward
    the binary 
    mass center and the jet precession is expected to be accelerated
    in the SMBHB models. After calculating the jet precession
    timescale and SMBHB evolution timescale in different models, we
    show that the acceleration of jet precession could reach 20 \% or
    even higher depending on the parameters of SMBHB systems and 
    accretion disks. Differential measurement of jet precession can be
    used to distinguish the 
    different precession models and determine the evolution timescale
    of SMBHB in galactic nuclei. With the measurement of precession
    acceleration, we could also determine the kinematic viscosity
    coefficient of accretion disk and the binary parameters.   

    Following the different physical mechanisms driving the evolution
    of SMBHB in galactic nuclei, in Section~\ref{sec:dynfr} we start
    our calculations of the hardening rate of SMBHB and the
    acceleration of jet precession with the regime when stellar
    dynamic friction affects the merger. 
    In Section~\ref{sec:massdisk}, we calculate the evolution of SMBHB
    and the time variation of jet precession because of the
    interaction of SMBHB with a massive circumbinary accretion disk,
    which is followed by the calculations for the scenario in which the
    evolution of SMBHB is dominated by the interaction between
    SMBHB and a non-massive circumbinary accretion disk in
    Section~\ref{sec:nmassdisk}. In Section~\ref{sec:gw}, we
    estimate the acceleration of jet precession because of the rapid  
    evolution of SMBHB dominated by gravitational wave radiations. 
    As an example, in section~\ref{sec:ngc} we discuss the differential
    observations of jet precession in a recent merged system, NGC1275
    (3C84), and the implications to the SMBHB in the object.
    Our discussions and conclusions on the results are given in
    Section~\ref{sec:dis}.


\section{Hardening of SMBHB because of galactic dynamic
  friction} 
\label{sec:dynfr}

\subsection{Unbound SMBHBs}

    Two SMBHs in merging galaxies are unbound until the gravitational
    force between the two SMBHs dominates the orbital motion when the
    separation of SMBHB $a$ is  
\begin{equation}
  a > a_{\rm H} = {G (M + m) \over \sigma^2} \simeq 1.12\times 10^6
    r_{\rm G} (1+q) \left({\sigma \over 200 {\rm Km/s}}\right)^{-2}
\end{equation}
    \citep{yu02}, where $\sigma$
    is the one-dimensional velocity dispersions of the primary galaxy,
    $r_{\rm G} = 2 G M/c^2$ is the Schwarzschild radius of the primary
    SMBH of mass $M$, and $q = m/M$ is the mass ratio of the secondary
    (of mass $m$) and the primary SMBHs. For $a > a_{\rm H}$, the
    evolution of SMBHB is dominated by galactic dynamic friction and
    the evolution timescale of $\tau_{\rm a}$ is approximately
    proportional to the separation $a$
\begin{equation}
  \tau_{\rm a}  = - {a \over (d a / dt)} \approx  \tau_{\rm H}
  \left({a \over a_{\rm H}}\right) ,
\end{equation}
    where $\tau_{\rm H} = - a_{\rm H} / v_{\rm dy}$ is the dynamic
    timescale at $a_{\rm H}$.   A negative sign for the definition of
    $\tau_{\rm a}$ is used because a SMBHB at galactic center never
    gets softer \citep{quinlan96}. The hardening rate of SMBHB due to
    galactic dynamic friction $v_{\rm dy}$ is approximately 
\begin{equation}
  v_{\rm dy} \approx -0.151 {\sigma_{\rm g}^3/\sigma^2} \ln\Lambda 
\label{eq:dyvel}
\end{equation} 
    \citep[e.g.][]{merritt00}.
    Here $\ln \Lambda \approx 2$ is the Column logarithm and
    $\sigma_{\rm g}$ is the one-dimensional velocity dispersion of the
    smaller (secondary) galaxy. Applying the empirical
    relation of central black hole mass $M$ and the stellar velocity
    dispersion $\sigma$ of host galaxy \citep{tremaine02}
\begin{equation}
  \log (M/M_\odot) = 8.13 + 4.02 \lg (\sigma / 200 {\rm Km / s}) 
  \label{eq:qcr}
\end{equation}
    to both the primary and the secondary galaxies, we obtain
    $\tau_{\rm H} \simeq 8.71 \times 10^5 M_8^{1.02/4.02} 
    q^{-3/4.02} (1+q) \, {\rm yr}$ and the SMBHB evolution
    timescale 
\begin{equation}
  \tau_{\rm a}  = - {a \over v_{\rm dy}} 
  \simeq  3.73 \times 10^7 M_8^{3.02/4.02} q_{-1}^{-3/4.02} \left({a
  \over 10^7 r_{\rm  G}}\right) \, {\rm yr} ,
\label{eq:dytu}
\end{equation}
   where $q_{\rm -1} = q/0.1$ and $M=M_8 \times 10^8
    M_\odot$. 

   When $a > a_{\rm H}$, jets, if present, would precess because of
   the orbital motion of the emitting black hole with the orbital
   period 
\begin{equation}
  P_{\rm orb} = 2 \pi \left[{a^3 \over G (M + m + M_*)}\right]^{1/2}
  \simeq 3.93 \times 10^6 M_8 \left({a \over 10^7 r_{\rm
      G}}\right)^{3/2} \left[{ 5 \over \left(1+q+{M_* \over
      M}\right)}\right]^{1/2} \, {\rm yr} ,
\label{eq:orb}
\end{equation}
    where $M_*$ is the mass of the stellar objects inside SMBHB
    orbit and $M_* > M+m$. Here a typical mass  $ M_* \sim 5 M$
    is used, because we are interested in a SMBHB with $a\ga a_{\rm H}$. 
    This is the shortest precession period in a SMBHB system with a
    given binary separation $a$ and has been introduced to explain the
    helical jet morphology on pc-scale and periodic optical outbursts
    observed in some blazars \citep[e.g.][]{villata99} and jet
    precession at Kpc scale or larger in some AGNs
    \citep[e.g.][]{wirth82}. From Equation~(\ref{eq:orb}), we can
    obtain the acceleration of jet precession in binary orbital motion
    because of the hardening of SMBHB 
\begin{equation}
  {d P_{\rm orb} \over d t} = -{3 \over 2} {P_{\rm orb} \over
  \tau_{\rm a}} .
\label{eq:accorb}
\end{equation}  

   From equations~(\ref{eq:orb}), (\ref{eq:dytu}), and
   (\ref{eq:accorb}), we get the acceleration of jet precession
   because of binary evolution
\begin{equation}
  {d P_{\rm orb} \over d t} \approx -0.16 \left({a \over 10^7 r_{\rm
      G}}\right)^{1/2} M_8^{1/4.02} q_{-1}^{3/4.02} \left({5 \over
      1+q+{M_* \over M}}\right)^{1/2} .
  \label{eq:acorbdy}
\end{equation} 
   As the precession period is within the observable range of jet 
   precession in the literature, 
   equation~(\ref{eq:acorbdy}) implies that one can measure the
   evolution of SMBHB owing to dynamic friction by detecting the  
   acceleration of jet precession.

\subsection{Bound SMBHBs}

   When $a < a_{\rm H}$, a SMBHB becomes bound, while a bound SMBHB
   becomes hard at a separation  
\begin{eqnarray}
  a_{\rm h} & = & {G mM \over 4 \sigma^2 (m+M)} \nonumber \\
  & = & 3.260 \times 10^4 r_{\rm G} M_8^{-1/2.01} {q_{-1} \over 1+ q} 
  \label{eq:ahd}
\end{eqnarray}
    \citep{quinlan96}. The evolution of SMBHB with separation $a_{\rm
      h} < a \la a_{\rm H}$ is still dominated by galactic dynamic
    friction, but the hardening timescale is approximated with
      $\tau_{\rm 
    dy} \propto a^{\gamma - 0.5}$ \citep{yu02}, where $\gamma$ is
    a fitting parameter in the Nuker law for the inner surface
    brightness profiles of galaxies
\begin{equation}
  I(r) = 2^{\beta - \gamma \over \eta} I_{\rm b} \left({r \over
    r_{\rm b}}\right)^{-\gamma} \left[1 + \left({r \over
      r_{\rm b}}\right)^{\eta}\right]^{-{\beta - \gamma \over \eta}} 
\end{equation}
    and $\eta$, $\beta$, $I_{\rm b}$, and $r_{\rm b}$ are fitting
    parameters, too. The break radius $r_{\rm b}$ is the point of
    maximum curvature in log-log coordinates and $\gamma$ is the
    asymptotic logarithmic slope inside $r_{\rm b}$. For core
    galaxies, $\gamma \la 0.3$, while for power law galaxies $\gamma
    \ga 0.5$. Therefore, the evolution timescale of a bound SMBHB is 
    approximately  
\begin{equation}
  \tau_{\rm a} \approx \tau_{\rm H} \left({a \over a_{\rm
      H}}\right)^{\gamma -0.5} .
  \label{eq:dytime}
\end{equation}
    This equation gives
\begin{eqnarray}
  \tau_{\rm a} & \simeq & 1.75 \times 10^7 \times 13.03^{-\gamma} 
  M_8^{(\gamma + 0.01)/2.01} q_{-1}^{-3/4.02} \nonumber \\
  && (1+q)^{1.5-\gamma}
  \left({a \over 10^5 r_{\rm G}}\right)^{\gamma - 0.5}  \, {\rm yr} 
\label{eq:dytf}
\end{eqnarray}
   for $a_{\rm h} < a \la a_{\rm H}$. Equation~(\ref{eq:dytime})
   implies that the evolution timescale of SMBHB decreases with binary
   separation for core galaxies but slowly increases for power law
   galaxies.  

   If the orbital motion and the angular momentum is dominated by
   the total mass of SMBHB, jet may precess because of the geodetic
   precession of the spin axis of the primary black hole about the
   total angular momentum \citep{begelman80}, of the orbital motion
   of the emitting black hole, and of the accretion disk precession
   due to tidal force of the inclined secondary SMBH outside the disk
   \citep{katz97}. If the rotating primary SMBH is misaligned with the
   binary total 
    angular momentum, the spin axis of the primary black hole
    undergoes geodetic precession about the total angular momentum
    with a period  
\begin{equation}
  P_{\rm geo} \simeq 2.6\times 10^7  M_8 q_{-1}^{-1}
  \left({a \over 10^4 r_{\rm G}}\right)^{5/2} \, {\rm yr}
\label{eq:precbbr}
\end{equation}
    \citep{begelman80,roos88}. This equation gives an 
    acceleration of jet precession due to the hardening of SMBHB 
\begin{equation}
    {d P_{\rm geo} \over dt} = - {5 \over 2} {P_{\rm geo} \over
    \tau_{\rm a}} .
\label{eq:dergeo1}
\end{equation}

   Equations~(\ref{eq:precbbr}), (\ref{eq:dergeo1}), and
   (\ref{eq:dytf}) shows that
\begin{equation}
  { d P_{\rm geo} \over dt} \simeq -1.2\times 10^3 13.03^\gamma
  M_8^{(2-\gamma)/2.01} q_{-1}^{-1.02/4.02} (1+q)^{\gamma-1.5}
  \left({a \over 10^5 r_{\rm G}}\right)^{3-\gamma} 
\label{eq:timdrgeo}
\end{equation}
   for $a_{\rm h} < a \la a_{\rm  H}$, implying that the acceleration
   of jet precession in the geodetic precession is too fast to be
   detected. 
   
    If the orbital plane of a SMBHB is inclined with respect to an
    accretion disk of radius $R_{\rm d}$ inside the binary
    orbit, the disk precesses like a rigid body owing to the tidal
    force of the secondary with a precession period  
\begin{equation}
  P_{\rm td} \simeq 5.1 \times 10^5 \, {\rm yr} \, M_8^2
  \left({a \over 10^5 r_{\rm G}}\right)^3 \left({R_{\rm d} \over
    10^4 r_{\rm G}}\right)^{-3/2} {(1+q)^{1/2} \over q_{-1}
    \cos\theta} 
\label{eq:precstell0}
\end{equation}
    \citep{katz97}, 
    where $\theta$ is the tilt angle of the disk plane and the binary
    orbital angular momentum. The disk size $R_{\rm d}$ could be the
    total radius extent of an accretion disk or the Bardeen-Peterson
    radius \citep{bardeen75,natarajan98}. When the SMBHB
    becomes hardening, equation~(\ref{eq:precstell0}) suggests that   
    the jet precession will be accelerated with 
\begin{equation}
    {d P_{\rm td} \over dt} = - 3 {P_{\rm td} \over \tau_{\rm a}} .
\label{eq:derst1}
\end{equation}
    To obtain equation~(\ref{eq:derst1}), we have assumed that the
    change of the disk radius $R_{\rm d}$ is insignificant, comparing
    to the variation of the binary separation. From
    equations~(\ref{eq:precstell0}), 
   (\ref{eq:derst1}), and (\ref{eq:dytf}), the jet precession because
    of binary-disk tidal interaction is accelerated with  
\begin{eqnarray}
  {d P_{\rm td} \over dt} & = & -0.087 \times 13^\gamma
  M_8^{(4.01-\gamma)/2.01} q_{-1}^{-1.02/4.02} \left(1 +
  q\right)^{\gamma-1} \nonumber \\
  && \left({a \over 10^5 r_{\rm G}}\right)^{3.5-
  \gamma} \left({R_{\rm d} \over 10^4 r_{\rm G}}\right)^{-3/2}
  \cos^{-1}\theta .
\end{eqnarray}
   The acceleration $d P_{\rm td} /dt$ of jet precession is very
   significant for power law galaxies but moderate for core galaxies. 

   From equations (\ref{eq:orb}),  (\ref{eq:accorb}), and 
   (\ref{eq:dytf}) the jet precession period due to binary orbital
   motion will change with time   
\begin{equation}
  {d P_{\rm orb} \over dt} \simeq -  7.5\times 10^{-4}
  M_8^{(2-\gamma)/2.01} q_{-1}^{3/4.02} (1+q)^{-2 +\gamma} \left({a
  \over 10^5 r_{\rm G}}\right)^{2 - \gamma} ,
\label{eq:nontimorb}
\end{equation}
    where we have taken $M_* \ll M + m$. Equation~(\ref{eq:nontimorb})
    shows that the change may be too small to be detectable.

\section{Evolution of SMBHB because of interaction with massive disk} 
\label{sec:massdisk}

   When a SMBHB becomes hard at $a \simeq a_{\rm h}$, the 
   evolution timescale $\tau_{\rm a}$ may be larger than the Hubble
   time and the binary may stall, if three-body interaction between
   SMBHB and stellar objects dominates the binary evolution
   \citep{quinlan96,yu02}. For a SMBHB stalling at the hard radius
   $a_{\rm h}$, equations~(\ref{eq:precbbr}), (\ref{eq:precstell0}),
   and (\ref{eq:orb}) together with equation~(\ref{eq:ahd}) give,
   respectively, the constant precession timescale 
\begin{eqnarray}
  P_{\rm geo} & \simeq & 4.99\times 10^8 M_8^{-0.98/4.02} q_{-1}^{3/2}
  (1+q)^{-5/2} \, {\rm yr} ,  
  \label{eq:pthdgeo} \\
  P_{\rm td} &\simeq & 1.77 \times 10^4 M_8^{1.02/2.01} {q_{-1}^2
  \over (1+q)^{5/2}} \left({R_{\rm d} \over 10^4 r_{\rm
  G}}\right)^{-3/2} {1 \over \cos\theta} \, {\rm yr} ,  
  \label{eq:pthdtd} \\
  P_{\rm orb} & \simeq & 1.63 \times 10^3 M_8^{1.02/4.02}
  {q_{-1}^{3/2} \over (1+q)^2} \, {\rm yr} .
  \label{eq:pthdorb}
\end{eqnarray}

    However, gas disk exists at the central region of AGNs, which
   should interact with SMBHB. In the AGN unification model, 
   the size of accretion disk around central SMBH is
   order of $10^4 r_{\rm G}$ while the broad emission line region and
   thick dust torus outside the accretion disk can be as large as
   $\sim 10^6 r_{\rm G}\sim 10 \, {\rm pc}$. Because the dust torus is
   geometrically thick and massive, the interaction between the
   secondary black hole and the dust torus is linear and the secondary
   cannot open a gap, probably leading to a rapid type I migration of
   the secondary toward the mass center \citep[e.g.][]{papal06}. If the
   accretion disk is geometrically thin and coplanar with the binary
   orbital plane, the secondary SMBH with mass ratio $q > q_{\rm min} = 
   {81 \pi \over 8} \alpha \delta^2 \simeq 3.2 \times 10^{-4}
   \alpha_{-1} (\delta / 0.01)^2$ will open a
   gap in the accretion disk and exchanges angular momentum with disk
   gas via non-linear Lindblad resonant binary-disk interaction
   \citep{lin86,armitage02}. Here, the viscous parameter $\alpha = 0.1 
   \alpha_{-1}$ is defined with the shear viscosity in r-$\phi$ plane,
   $\nu_1 = \alpha c_{\rm s} H$ with $H$ the scale height of the
   unperturbed accretion  disk and $c_{\rm s}$ the sound speed, and
   $\delta = H/r$ is the disk openning angle at radius $r$. 
   The migration of the secondary black hole is called Type II
   migration. If the disk mass inside the binary orbit is larger than
   the mass of the secondary SMBH, the migration timescale
   of the secondary SMBH is the disk viscous timescale
   \citep{lin86,armitage02,papal06}. If the circumbinary disk is
   massive, the secondary SMBH migrate also on a disk viscous
   timescale even if the orbital plane and the accretion disk is
   misaligned \citep{ivanov98}. Because the mass ratio of a
   SMBHB formed in galaxy mergers within Hubble time is $q \ga
   10^{-3} > q_{\rm min}$ \citep{yu02,liu04}, we consider only type II
   migration. 

   For a type II migration, the secondary black hole  
   migrates inwards on a viscous timescale 
\begin{equation}
  \tau_{\rm a} \approx t_\nu \simeq - {a \over v_{\rm r}} \simeq
  {2 \over 3} { a^2 \over \nu_1} .
  \label{eq:davis}
\end{equation}
    From equations~(\ref{eq:davis}) and $\nu_1 = \alpha c_{\rm s} H$,
    we have 
\begin{equation}
  \tau_{\rm a} \simeq 1.66 \times 10^6 M_8 \alpha_{-1}^{-1}
    \delta_{-2}^{-2} \left({a \over 10^4 r_{\rm
    G}}\right)^{5/4} \, {\rm yr} 
  \label{eq:davis1}
\end{equation}
    where $\alpha_{-1} = \alpha/0.1$ and $\delta_{-2} = \delta_0
    /0.01$. Here for convenience, we have written 
\begin{equation}
  \delta \equiv \delta_0 \left({r \over 10^3 r_{\rm
      G}}\right)^{\lambda} ,
  \label{eq:defopen}
\end{equation}
    where $\delta_0$ is the disk opening angle at $r= 10^3 r_{\rm G}$
    and weakly depends on $\alpha$, the central black hole mass $M$,
    and the accretion rate $\dot{m}$. For a standard $\alpha$-disk,
    $\lambda = 1/8$ if the disk is gas pressure and
    free-free absorption dominated, while $\lambda = 1/20$ if
    the disk is gas pressure and electron scattering dominated
    \citep{kato98}. To get equation~(\ref{eq:davis1}), we have assumed
    that the disk is gas pressure and free-free absorption dominated
    with $\lambda = 1/8$, which would be valid for $r \ga 2.6 \times
    10^3 r_{\rm G} \dot{m}_{-1}^{2/3}$ \citep{kato98}, $\dot{m} =
    \dot{M} / \dot{M}_{\rm Edd} = 0.1 \dot{m}_{-1}$ is the
    dimensionless accretion rate, and the Eddington accretion rate
    $\dot{M}_{\rm Edd} = L_{\rm Edd} / 0.1 c^2$ is related to the
    Eddington luminosity $L_{\rm Edd} = 1.26\times 10^{46} M_8 \, {\rm
    erg \; s^{-1}}$.

    Because the disk is massive, the total angular momentum of the 
    binary-disk system is dominated by the disk mass and there is no 
    geodetic jet precession around binary orbital angular
    momentum. However, the inner disk region misaligned with a
    rotating central SMBH may be warped and become aligned due to
    Bardeen-Peterson effect \citep{bardeen75}, jets may precess
    because of the tidal interaction of a misaligned secondary black
    hole and the warped inner disk. From
    equation~(\ref{eq:precstell0}), (\ref{eq:derst1}), and
    (\ref{eq:davis1}), we have 
\begin{equation}
  {d P_{\rm td} \over dt} \simeq -2.61 M_8 \alpha_{-1} \delta_{-2}^2
  \left({a \over 10^4 r_{\rm G}}\right)^{7/4} \left({r_{\rm BP} \over
  50 r_{\rm G}}\right)^{-3/2} \left(1 + q\right)^{1/2} q_{-1}^{-1}
  \cos^{-1}\theta ,
\label{eq:acctdev}
\end{equation}
   where $r_{\rm BP}$ with typical value of order $\sim 50 r_{\rm G}$ 
   is the Bardeen-Peterson radius out to which the accretion disk flow
   is aligned with the black hole spin axis
   \citep{bardeen75,natarajan98}. Equation~(\ref{eq:acctdev}) implies 
   that the acceleration of the jet precession is significant and can
   be detected very easily. 

   Jets will also precess because of the orbital motion of the
   emitting primary black hole in the case of massive circumbinary
   accretion disk. However, the acceleration of jet precession due to 
   binary orbital motion 
\begin{equation}
 {d P_{\rm orb} \over dt} = -{3 \over 2} {P_{\rm
   orb} \over \tau_{\rm a}} \simeq - 2.5 \times 10^{-4} \alpha_{-1}
   \delta_{-2}^2 \left({a \over 10^4 r_{\rm G}}\right)^{1/4}
   (1+q)^{-1/2}
\end{equation}
   may be too small to be detectable. 

   When the inner disk region becomes aligned with the rotating black
   hole but misaligned with the outer inclined accretion disk, the
   misaligned disk region and the spin axis of central rotating black
   hole would precess with a precession timescale 
\begin{equation}
  P_{\rm BP} \simeq  1.51 \times 10^6  a_*^{11/16}
  \alpha_{-1}^{13/8} \dot{m}_{-1}^{-7/8} M_8^{-1/16} \, {\rm yr} 
  \label{eq:at1}
\end{equation}
    \citep{natarajan98}, where $a_*$ is the spin parameter of the
    primary SMBH. Equation~(\ref{eq:at1}) shows that the jet precession
    due to Bardeen-Peterson effect is
    independent of the evolution of SMBHB and does not change with time 
    on the timescale that we are interested.

\section{Evolution of SMBHB because of interaction with a non-massive
  disk} 
\label{sec:nmassdisk}

   When the secondary black hole migrate inwards on a viscous
   timescale and reaches a critical radius $r_{\rm m}$, the disk mass
   inside the binary orbit will equal to the mass of the secondary
   black hole. In a gas-pressure and electron-scattering dominated
   $\alpha$-disk, the unperturbed disk surface density is 
\begin{equation}
  \Sigma \simeq 2.4 \times 10^5 \alpha_{-1}^{-4/5} M_8^{1/5}
  \dot{m}_{-1}^{3/5} r_3^{-3/5} \, {\rm g\; cm^{-2}} 
  \label{eq:sigma}
\end{equation}
    \citep{kato98}, where $r_3 = r/ 10^3 r_{\rm G}$. Note that we have 
    used different $\alpha$ prescription. From
    equation~(\ref{eq:sigma}), we can estimate the disk mass $M_{\rm 
    d}$ inside radius $r$ 
\begin{equation}
  M_{\rm d} \simeq {10 \over 7} \pi \Sigma r^2 \simeq 4.9 \times 10^5
  \alpha_{-1}^{-4/5} M_8^{11/5} \dot{m}_{-1}^{3/5} r_3^{7/5} M_\odot .
  \label{eq:dmass}
\end{equation}
    When $M_{\rm d} = m$, from equation~(\ref{eq:dmass}) we have 
\begin{equation}
  r_{\rm m} = 8.6 \times 10^3 q_{-1}^{5/7} \alpha_{-1}^{4/7}
   \dot{m}_{-1}^{-3/7} M_8^{-6/7} r_{\rm G} .
\label{eq:rm}
\end{equation}
   When $a <  r_{\rm m}$, the disk mass $M_{\rm d}$ inside the binary 
   orbit is smaller than the mass of the secondary and the migration
   of the secondary will be reduced \citep{syer95,ivanov99}. 
   If the disk is gas-pressure and electron-scattering dominated, 
   the migration timescale is approximately  
\begin{equation}
  \tau_{\rm a}  \simeq  \left({152 \over 112}\right) \left({15 \over 
    152}\right)^{5/19} \left({16 \over 11}\right)^{16/19}
  \left({M_{\rm d} \over m}\right)^{-14/19} t_\nu 
\label{eq:migtimred}
\end{equation}
   \citep{ivanov99}, 
   where $M_{\rm d} = \dot{M} t_\nu$ is disk mass inside the binary
   orbit and $t_\nu$ is the viscous timescale of unperturbed accretion
   disk at $r = a$. Taking $t_\nu = (2/3) a^2/\nu_1$ and from
   equations~(\ref{eq:defopen}) and (\ref{eq:migtimred}), we have 
\begin{equation}
  \tau_{\rm a} \simeq 8.95 \times 10^6 \dot{m}_{-1}^{-14/19}
  q_{-1}^{14/19} M_8^{5/19} \alpha_{-1}^{-5/19} \delta_{-2}^{-10/19} 
  \left({a \over 10^3 r_{\rm G}}\right)^{7/19} \, {\rm yr}  
\label{eq:migtimred1}
\end{equation}
   for a gas-pressure and electron-scattering dominated 
   disk with $\lambda = 1/20$. 

   If the rotating primary black hole is inclined to the binary total
   angular momentum, its spin axis will precess geodeticaly. From
   equations~(\ref{eq:precbbr}), (\ref{eq:dergeo1}), and
   (\ref{eq:migtimred1}), we obtain the acceleration of geodetic
   precession
\begin{equation}
  {d P_{\rm geo} \over dt} \simeq - 2.3 \times 10^{-2}
  \dot{m}_{-1}^{14/19} M_8^{14/19} q_{-1}^{-33/19} \alpha_{-1}^{5/19}
  \delta_{-2}^{10/19} \left({a \over 10^3 r_{\rm G}}\right)^{81/38} .
\label{eq:geocmbl}
\end{equation}
   
   When the disk mass inside the binary orbit is less than the mass of
   the secondary black hole, the secondary will warp, twist, and
   realign the inner accretion disk on a short timescale
   \citep{ivanov99}. However, the realignment will stop at the
   Bardeen-Peterson radius $r_{\rm BP}$ and the disk region at $r <
   r_{\rm BP}$ will remain aligned with rotating primary black hole
   and misaligned with the binary orbital plane \citep{liu04}. So, the
   inner accretion disk has $r < r_{\rm BP}$ and thus the jet
   orientation precess due to the tidal interaction of the secondary
   black hole. Equations~(\ref{eq:precstell0}), (\ref{eq:derst1}), and
   (\ref{eq:migtimred1}) give the acceleration of the precession
   period because of binary-disk tidal interaction
\begin{equation}
  {d P_{\rm td} \over dt} \simeq - 4.8\times 10^{-4} \dot{m}_{-1}^{14/19}
  M_8^{33/19} q_{-1}^{-33/19} \alpha_{-1}^{5/19} \delta_{-2}^{10/19}
  \left({a \over 10^3 r_{\rm G}}\right)^{50/19} \left({r_{\rm BP} \over
  50 r_{\rm G}}\right)^{-3/2} {\left(1+q\right)^{1/2} \over \cos\theta} .
\label{eq:tdcml}
\end{equation}
   The time variation of jet precession is very small. 

   When the disk mass $M_{\rm d}$ within the binary orbit is less than
   that of the secondary black hole, a SMBHB with an inclined orbital
   plane warps and realigns the inner disk region outside its orbit to
   a typical transitional radius $r_{\rm al}$ \citep{ivanov99}. The
   transitional radius 
   $r_{\rm al}$ of the inner warped and the outer unperturbed disk
   regions depends on how warps communicate in the disk. Taking
   into account the internal hydrodynamics of the disk itself,
   \citet{papaloizou83} showed that for $\alpha > \delta = H/r$  
   warp transfers on a timescale $t_{\rm wp} \simeq 2 r^2 /3 \nu_2$,
   where $\nu_2$ is the vertical viscosity and relates to the shear
   viscosity $\nu_1$ in r-$\phi$ plane with $\nu_2 \approx \nu_1 f_0/(2
   \alpha^2)$ and $f_0 = (1 + 7\alpha^2) / (1 + \alpha^2/4)$
   \citep{kumar85,kumar90,ogilvie99}. 

   The quadrupole contribution of the secondary black hole to the 
   gravitational potential would causes the precession of the major
   axis of an elliptical orbit in the disk with frequency  
\begin{equation}
  \Omega_{\rm ap} = {3 \over 4} q \left({a \over r}\right)^2
  \Omega_{\rm K} 
  \label{eq:quprec}
\end{equation}
    \citep{ivanov99}, 
    where $\Omega_{\rm K}$ is the Keplerian angular velocity at
    $r$. The lines of nodes precesses with frequency $\Omega_{\rm np}
    = - \Omega_{\rm ap}$. \citet{liu04} showed that for $f_0 = 1$ the
    transitional radius $r_{\rm al}$ can be estimated by using $t_{\rm
    wp} \simeq \Omega_{\rm ap}^{-1}$, which gives 
\begin{equation}
  r_{\rm al} \simeq \left(q \alpha / f_0 \right)^{1/2} \delta^{-1} a ,
  \label{eq:ral}
\end{equation}
    where $r_{\rm al} \la r_{\rm m}$. 
    The precession period $P_{\rm ap}$ of the aligned inner disk is
    determined by the precession of the lines of nodes at $r_{\rm al}$  
\begin{eqnarray}
  P_{\rm ap} & \simeq & 2 \pi \Omega_{\rm ap}^{-1}  \nonumber \\
 & = & \left({8 \sqrt{2} \over 3} \pi\right) 10^{21 \lambda 
    /2 (1 + \lambda)} \left({r_{\rm G} \over c}\right) \left({\alpha
    \over f_0 \delta_0^2}\right)^{7/4(1+\lambda)} \nonumber \\
  & & q^{(3-4\lambda)/4(1+\lambda)} \left({a \over r_{\rm G}}
 \right)^{(3-4\lambda)/2(1+\lambda)} .
  \label{eq:prec2}
\end{eqnarray}
    For a gas-pressure and electron-scattering dominated thin 
    disk, $\lambda=1/20$ and the precession period is
\begin{equation}
  P_{\rm ap}  \simeq 2.52\times 10^5 \, {\rm yr} \, M_8 q_{-1}^{2/3}
  \alpha_{-1}^{5/3} f_0^{-5/3} \delta_{-2}^{-10/3} \left({ a \over 10^3
  r_{\rm G}}\right)^{4/3} .
  \label{eq:prec4}
\end{equation}
    Because the warp transfer timescale $t_{\rm wp} = (2 \alpha / f_0
    ) t_v$ is much smaller than the viscous timescale $t_v$ for a
    standard thin disk with $\alpha \ll 1$, an assumption of
    rigid-body like disk precession is reasonable. If the primary
    SMBH is rotating and misaligned with the binary orbit plane, 
    warp transfers quickly inwards and stall at the Bardeen-Peterson
    radius $r_{\rm BP}$ \citep{bardeen75}. The 
    disk region within $r_{\rm BP}$ and the spin axis of the
    rotating primary BH would also precess with a timescale $P_{\rm
    BP}$. 

    From equation~(\ref{eq:prec2}), the acceleration of the precession
    of a warped circumbinary disk is
\begin{equation}
  {d\ln P_{\rm ap} \over d t}   =  {d\ln M \over dt} - 
   {7 \over 2 (1 + \lambda)} {d\ln \delta_0 \over dt} + 
  {3 - 4\lambda \over 2 (1+\lambda)} {d\ln a \over dt} 
\label{eq:derap1}
\end{equation}
   where we have assumed a constant mass ratio $q$ during
   the evolution of SMBHB. In accretion disk theory, the opening
   angle $\delta_0$ depends on accretion rate and the mass of central
   black hole 
\begin{equation}
  \delta_0 \propto \dot{m}^\mu M^\zeta 
\label{eq:delta0}
\end{equation}
    with $\mu > 0$ and $\zeta>0$ \citep{kato98}. Substituting
    equation~(\ref{eq:delta0}) into equation~(\ref{eq:derap1}), we
    have
\begin{equation}
  {d P_{\rm ap} \over d t}  \simeq {7 \mu \over
    2 (1 + \lambda)} {P_{\rm ap} \over \tau_{\dot{m}}} - 
    {3 - 4\lambda
    \over 2 (1+\lambda)} {P_{\rm ap} \over \tau_{\rm a}} , 
  \label{eq:derap2}
\end{equation}
    where $\tau_{\dot{m}} = - \dot{m} / (d \dot{m} / dt)$ is the
    variation time scale of accretion rate, which is equivalent to the
    typical lifetime of the AGN and may be determined by the
    environment, for example, the supply of the gas from the 
    galactic disk to the accretion disk and the interaction of
    accretion disk and the stellar objects passing through the disk.
    To obtain equation~(\ref{eq:derap2}), we have assumed that the
    mass of the primary SMBH is insignificant on the timescale
    which we are interested in here, implying that the variation
    timescale of the black hole mass $\tau_M = M / (dM/dt)$ is much
    longer than the binary hardening timescale $\tau_{\rm a}$. 
    Equation~(\ref{eq:derap2}) suggests that the decrease of accretion
    rate decelerates the jet precession but the hardening of SMBHB
    accelerates it. If a SMBHB is long-lived and passes through the
    active phase of a galaxy, namely $\tau_{\dot{m}} \ll \tau_{\rm a}$,
    equation~(\ref{eq:derap2}) gives a deceleration rate of jet
    precession
\begin{equation}
  {d P_{\rm ap} \over d t} \simeq {2 \over 3 } 
    {P_{\rm ap} \over \tau_{\dot{m}}} 
\label{eq:derapm}
\end{equation}
    for a gas-pressure and electron-scattering dominated standard thin
    disk with $\mu = 1/5$ and $\lambda =
    1/20$. Equation~(\ref{eq:derapm}) suggests that from the
    measurement of jet precession and its deceleration rate, we can 
    determine the disk evolution and the life time of an
    individual radio source. For a short-lived SMBHB
    with $\tau_{\rm a} \ll \tau_{\dot{m}}$, jet precession is
    accelerated 
\begin{equation}
  {d P_{\rm ap} \over d t} \simeq - {4 \over 3} { P_{\rm ap} \over
 \tau_{\rm a}} 
  \label{eq:derapb}
\end{equation}
    for a gas-pressure and electron-scattering dominated thin disk.

    From equations~(\ref{eq:derapb}) and (\ref{eq:migtimred1}), we
    obtain the acceleration of the precession of a warped disk due to
    the reduced migration of the secondary black hole 
\begin{equation}
  {d P_{\rm ap} \over dt} \simeq - 3.8\times 10^{-2}
  \dot{m}_{-1}^{14/19} M_8^{14/19} q_{-1}^{-4/57} f_0^{-5/3}
  \alpha_{-1}^{110/57} \delta_{-2}^{-160/57} \left({a \over 10^3
  r_{\rm G}}\right)^{55/57} .
\label{eq:apcml}
\end{equation}

   At $a\sim 10^3 r_{\rm G}$, the orbital period of a SMBHB is $P_{\rm
     orb} \simeq 8.8 \left({a \over 10^3 r_{\rm G}}\right)^{3/2} {M_8
     \over (1+q)^{1/2}} \, {\rm yr}$ and the jet precession because of
   the binary orbital motion would be nearly constant as $d P_{\rm
     orb} / dt \simeq - {3 \over 2} { P_{\rm orb} \over 
   \tau_{\rm a}} \la 10^{-5}$.

\section{Rapid Evolution of SMBHB owing to gravitational wave
     radiation} 
\label{sec:gw}

   When $a$ is order of $10^2 r_{\rm G}$, the loss of the orbital
   angular momentum because of gravitational wave radiation becomes
   important \citep{armitage02} and the in-spiraling velocity of the
   secondary black hole due to gravitational wave radiation is 
\begin{eqnarray}
  \dot{a}_{\rm gw} & = & - {64 G^3 M^3 q \left(1 + q\right) \over 5
    c^5 a^3 } f = - {8\over 5} \left({r_{\rm G} \over a}\right)^3 q 
  \left(1 + q\right) f c \cr
  & \simeq & -4.8 \times 10^3 \left({a \over
    10^3 r_{\rm G}}\right)^{-3} f q_{-1} \left(1 + q\right) \, {\rm cm
    \; s^{-1}}
  \label{eq:agw}
\end{eqnarray}
   \citep{peters63}, where $f$ is a function of eccentricity $e$
\begin{equation}
  f = \left(1 + {73 \over 24} e^2 + {37 \over 96} e^4\right) 
  \left(1 - e^2\right)^{-7/2} .
\label{eq:fec}
\end{equation}
   At large $a$, the evolutions of the SMBHB and the accretion
   disk are coupled and the migration timescale of the secondary SMBH
   is given by equation~(\ref{eq:migtimred1}). When $a$ is small, the
   loss of the binary orbital angular momentum is dominated by the
   gravitational wave radiation and the hardening timescale due to the
   gravitational wave radiation is given with 
\begin{equation}
  \tau_{\rm a} = -a/\dot{a}_{\rm gw} \simeq 1.95 \times 10^4 M_8
  q_{-1}^{-1} \left(1+q\right)^{-1} f^{-1} \left({a \over 10^2 r_{\rm
  G}}\right)^4 \, {\rm yr} .
\label{eq:timgw}
\end{equation} 
   At a critical radius $a = a_{\rm gw}$, the in-spiraling timescale
   due to the gravitational wave radiation, $\tau_{\rm gw}$,
   approximately equals to the migration timescale of the secondary
   SMBH because of the 
   interaction between the SMBHB and a non-massive accretion disk, 
   $\tau_{\rm a}$. From equation~(\ref{eq:migtimred1}) and $\tau_{\rm
   gw} = - a_{\rm gw} / \dot{a}_{\rm gw}$, we obtain 
\begin{equation}
  a_{\rm gw} \simeq 4.27 \times 10^2 r_{\rm G} \dot{m}_{-1}^{-14/69}
  M_8^{-14/69} q_{-1}^{11/23} \left(1+q\right)^{19/69}
  \alpha_{-1}^{-5/69} \delta_{-2}^{-10/69} f^{19/69} .
\label{eq:radgw}
\end{equation}

    From equations~(\ref{eq:precbbr}), (\ref{eq:dergeo1}), and 
    (\ref{eq:timgw}), we obtain the acceleration of jet precession
    because of the geodetic precession
\begin{equation}
  {d P_{\rm geo} \over dt} \simeq 3.3 \times 10^{-2} \left(1 +
  q\right) f \left({a \over 10^2 r_{\rm G}}\right)^{-3/2} .
\label{eq:atgwg}
\end{equation}
   Equation~(\ref{eq:atgwg}) implies that the time variation of jet
   geodetic precession depends only on the binary separation and is 
   independent of the parameters of accretion disk and the SMBHB. 
   
   From equations~(\ref{eq:orb}), (\ref{eq:accorb}),
   (\ref{eq:precstell0}), (\ref{eq:derst1}), and (\ref{eq:timgw}), 
   the accelerations of the jet precession owing to the tidal
   interaction of binary-disk and to the orbital motion are,
   respectively, 
\begin{eqnarray}
  {d P_{\rm td} \over dt} &\simeq & 2.2 \times 10^{-4} M_8 \left( 1+
  q\right)^{3/2} f \left({a \over 10^2 r_{\rm G}}\right)^{-1}
  \left({r_{\rm BP} \over 50 r_{\rm G}}\right)^{-3/2} \cos^{-1}\theta
  , \\
    {d P_{\rm orb} \over dt} &\simeq & 2.1 \times 10^{-5} q_{-1}
  \left(1+ q\right)^{1/2} f \left({a \over 10^2 r_{\rm
  G}}\right)^{-5/2} . 
\end{eqnarray}
   The acceleration are insignificant. 

   From equations~(\ref{eq:prec4}), (\ref{eq:derapb}), and
   (\ref{eq:timgw}), we get the acceleration of jet precession because
   of precession of a warped circumbinary disk  
\begin{equation}
  { d P_{\rm ap} \over dt} \simeq -0.80 q_{-1}^{5/3} (1+q) f
  \alpha_{-1}^{5/3} \delta_{-2}^{-10/3} f_0^{-5/3} 
  \left({a \over 10^2 r_{\rm G}}\right)^{-8/3} .
\label{eq:acgwap}
\end{equation}

\section[]{Differential jet precession and SMBHB in NGC1275}
\label{sec:ngc}

    In previous sections, we discussed the evolution of SMBHB in
    different driving regimes and the corresponding acceleration of
    jet precession. 
    Equations~(\ref{eq:dergeo1}), (\ref{eq:derst1}),
    (\ref{eq:accorb}), and (\ref{eq:derapb}) suggest that the
    acceleration of jet precession because of SMBHB hardening
    can be written integrally   
\begin{equation}
  {d P_{\rm pr} \over d t} \simeq - \Lambda {P_{\rm pr} \over
  \tau_{\rm a}} ,
\label{eq:decgen}
\end{equation}
    with ${4\over 3} \leq \Lambda \leq 3$. Our calculation show that
    the acceleration of jet precession because of SMBHB evolution is
    significant and could be measured on the timescale of jet
    precession period. If we measure the jet precession period $P_{\rm
    pr}$ and compute the acceleration rate ${d P_{\rm pr}/ dt}$, we
    can determine the evolution timescale of a SMBHB in galactic
    nuclei with 
\begin{equation}
  \tau_{\rm a} \simeq  - \Lambda {P_{\rm pr} \over \dot{P}_{\rm
      pr}} .
\label{eq:harden}
\end{equation}
    With these calculations, we will discuss, as an example, the
    differential jet precession in the radio galaxy NGC1275. 

\subsection{Jet precession with constant timescale}

    The FRI radio galaxy NGC1275 (3C84) is a recent merger system
    at redshift $z = 0.01756$ \citep[e.g.][]{holtzman92} and the mass of
    central SMBH is measured with molecular gas hydrodynamic method to
    be $M = 3.4\times 10^8 \, M_\odot$ \citep{wilman05}. The object
    has a bolometric luminosity $L_{\rm bol} \simeq
    1.07\times 10^{44} \, {\rm ergs\; s^{-1}}$ (with $H_0 = 75\, {\rm
    km \; s^{-1} \; Mpc^{-1}}$ and $q_0 = 0.5$) \citep{march04}. With 
    the measured black hole mass and the bolometric
    luminosity, we obtain the dimensionless accretion rate $\dot{m}
    \simeq 2.4\times 10^{-3} \epsilon_{-1}^{-1}$.  \citet{dunn06}
    imaged the S-symmetrical morphologies of jets and emission line
    structure and differentially measured jet precession timescale
    by identifying four components of different orientations
    in order of formation: ancient bubbles, ghost bubbles, outer
    lobes, and inner jets. The observations of precession angle 
    $\Delta{\phi}$, time difference $\Delta{t}$, the number of
    complete cycles $n$ between two components in succession,
    the observed precession period $P_{\rm pr}$ 
    are taken from \citet{dunn06} and summarized in
    Table~\ref{obser}. The number $n$ is the precession cycle number
    between two successive components and is estimated  
    with $P_{\rm pr} = 360\degr \Delta{t} / (\Delta\phi + n
    360\degr)$. The observations suggests that the activity of
    NGC1275 is intermittent and the precession timescale are
    significantly different for different episodic
    activities. Intermittence of activity and significant differences
    of jet precession periods for different activity episodes are 
    also observed in the Seyfert 1.5 galaxy Mrk 6 \citep{kharb06}. For
    both NGC1275 and Mrk6, the observations shows that the jet
    orientation at the beginning of each episodic activity are
    significantly different from that at the end of last episodic
    activity, implying that the spin axis of central black hole
    precesses even when the source is dormant or at very weak
    activity.

\begin{table}  
\caption{ \label{obser} Observations of jet
    precession in NGC1275 (3C84) by \citet{dunn06}. Both $\Delta t$
  and $P_{\rm pr}$ are in units of $10^7\, {\rm yr}$.}
\begin{tabular}{rr@{ $=$ }lr@{ $=$ }l}
 & \multicolumn{2}{c}{North} &\multicolumn{2}{c}{South} \\
\hline
Ancient$\to$Ghost&$\Delta{\phi}$&$61\degr$&$\Delta{\phi}$&$113\degr$\\
&$\Delta{t}$&$3.8$&$\Delta{t}$&$7.2$\\
$n=0$&$P_{\rm pr}$&$22.4$&$P_{\rm pr}$&$22.9$\\
$n=1,2$&$P_{\rm pr}$&$3.25$&$P_{\rm pr}$&$3.11$\\
\hline
Ghost$\to$Outer&$\Delta{\phi}$&$274\degr$&$\Delta{\phi}$&$293\degr$\\
&$\Delta{t}$&$6.1$&$\Delta{t}$&$6.2$\\
$n=0$&$P_{\rm pr}$&$8.01$&$P_{\rm pr}$&$7.62$\\
$n=1$&$P_{\rm pr}$&$3.46$&$P_{\rm pr}$&$3.42$\\
\hline
Outer$\to$Jet&$\Delta{\phi}$&$201\degr$&$\Delta{\phi}$&$249\degr$\\
&$\Delta{t}$&$1.5$&$\Delta{t}$&$1.9$\\
$n=0$&$P_{\rm pr}$&$2.68$&$P_{\rm pr}$&$2.75$\\
\hline
Ancient$\to$Outer\\
$n=0$&$\langle d P_{\rm pr}/dt\rangle $&$-2.55$\\ 
     &$\langle P_{\rm pr}\rangle$&$15.23$\\
$n=1,2$&$\langle P_{\rm pr}\rangle$&$3.31$\\
\hline
Ghost$\to$Jet\\
n=0&$\langle d P_{\rm pr}/dt\rangle$&$-1.30$\\
    &$\langle P_{\rm pr}\rangle$&$5.27$\\
n=1&$\langle d P_{\rm pr}/dt\rangle$&$-0.19$\\
    &$\langle P_{\rm pr}\rangle$&$3.08$\\
\hline
\end{tabular}
\end{table}

    \citet{dunn06} assumed that the jet precession
    in NGC1275 remains steady for many cycles and the observed
    differences of precession timescale are due to the missing of
    different precession cycles when no bubble detaches. With the
    assumption, the significantly different precession timescale
    from Ancient bubbles through Ghost bubbles to outer lobes
    are reconciled with one period $P_{\rm pr} \simeq (3.31
    \pm 0.46) \times 10^7 \, {\rm yr}$, but the currently active jets
    still precess with a significantly shorter timescale $P_{\rm pr}
    \simeq (2.72\pm 0.54)\times 10^7 \, {\rm yr}$ \citep{dunn06}. Here
    the errors have included the measurement error of precession
    timescale, $\sigma_P/P_{\rm pr} \sim 20 \%$. The cycle number is
    given in Table~\ref{obser}. The jet precession from Ghost bubbles
    to active jets is accelerated with $\langle d P_{\rm pr}/dt
    \rangle \simeq -0.19 $, which implies that the jet precession is
    due to SMBHB at center. Equation~(\ref{eq:harden}) gives a
    model-independent evolution timescale of SMBHB in NGC1275
    $\tau_{\rm ob} \simeq \left({\Lambda \over 2}\right) 3.62
    \times 10^8 \, {\rm yr}$. If we knows the 
    binary separation $a$, we can calculate a model-independent
    binary hardening rate or the migration velocity of the secondary 
    $v_{ob} \simeq - 54 \left({ a \over 0.2 \,
    {\rm pc}}\right) \left({\Lambda \over 2}\right)^{-1} \, {\rm
    cm/s}$ with $4/3 \leq \Lambda \leq 3$.

    Jets precess with constant timescale 
    through several duty cycles of activity
    implies that the precession in NGC1275 is independent of
    the accretion. All the models for jet 
    precession depending on accretion disk are excluded and the only 
    reasonable scenarios for the jet precession from one bubble to
    another are the geodetic precession or binary orbital motion. If the
    precession is due to the orbital motion of a SMBHB, the observed
    period and equation~(\ref{eq:orb}) gives $a_{\rm orb} \simeq 1.81
    \times 10^7 r_{\rm G} \left[\left(1 + q+ {M_* \over
    M}\right)/5\right]^{1/3} \simeq 5.9 \times 10^2 \, {\rm pc}$,
    which is much larger than the bound radius
    $a_{\rm H} \simeq 7.1\times 10^5 r_{\rm H}$ and implies an unbound 
    SMBHB in NGC1275. An unbound SMBHB is consistent with
    the observations of recent merger. However,
    equation~(\ref{eq:acorbdy}) suggests a variation of precession
    timescale from the ancient bubbles to outer lobes for the south
    components
\begin{equation}
  {\Delta{P}_{\rm pr} \over P_{\rm pr}} \simeq {d P_{\rm orb} \over
  dt} \left[{\Delta{t}_{\rm A\rightarrow G} + \Delta{t}_{\rm G\rightarrow
  O} \over 2 P_{\rm pr}}\right] 
  \simeq - 0.59 q_{-1}^{3/4.02} \left({5 \over 1 + q + {M_*
  \over M}}\right)^{1/2} ,
\label{eq:orbngc}
\end{equation}
    which together with the assumption of steady precession gives an
    upper limit $q \la 2 \times 10^{-2} \left[\left(1 + q + {M_* \over
    M}\right) / 5\right]^{2.01/3}$.  
    Equation(\ref{eq:dytu}) shows that to form such a binary in a
    minor merger, the dynamical friction timescale at $a \sim 20 {\rm 
    Kpc}$ is $\tau_{\rm a} \ga 1.4 \times 10^9 \left({1 + q + {M_*
    \over M} \over 10^3}\right)^{-1/2}\, {\rm yr}$, where $M_*$ is the 
    stellar mass within the binary orbit with $a\sim 20 {\rm Kpc}$. A
    minor merger with $q \la 2\times 10^{-2} $ is unlikely to be
    observable for such a long timescale. In this scenario, a steady 
    precession from the ancient bubbles to the outer lobes is also
    inconsistent with the acceleration of jet precession from the
    Ghost bubbles to the present active jets. 

    The second possible precession scenario independent of accretion
    rate is the geodetic precession of the primary SMBH. From
    equation~(\ref{eq:precbbr}), the constant precession timescale
    implies 
\begin{equation}
  a_{\rm geo} \simeq 6.7 \times 10^3 r_{\rm G} q_{-1}^{2/5} ,
\label{eq:geongc}
\end{equation}
   which is smaller than $a_{\rm h}$ for $q \ga 2.0 \times
   10^{-2}$.  When the source is at very weak 
   activity or dormant with an accretion rate much smaller than 
   its current accretion rate and also the typical accretion rate for FRI
   radio galaxies, namely $\dot{m} \ll 10^{-3}$, the accretion disk
   cannot be a standard thin disk but geometrically thick and
   optically thin advection dominated accretion flows (ADAFs)
   \citep{narayan94,meyer94,abra95}. The interaction of an ADAF and
   the secondary black hole in a binary system is dynamically
   negligible \citep{liu04}. For such a binary-disk system, the total
   angular momentum is dominated by the orbital angular momentum and
   the jet precession is geodetic. For a ADAF-binary system, the
   jet precession other than the geodetic precession is the orbital
   motion with period $P_{\rm orb} \simeq 5 \times 10^2 \, {\rm yr}$.
   For an accretion disk with accretion rate $\dot{m} = 2.4\times
   10^{-3}$, the inner region is ADAF and the outer part of the disk
   is a standard thin disk. The transition radius between the two
   different accretion modes is
\begin{equation}
  r_{\rm tr} \simeq 18.3 \dot{m}^{-0.85} r_{\rm G} \simeq 3.1 \times
  10^3 r_{\rm G} 
\label{eq:modetr}
\end{equation}
   \citep{meyer00,liu02b}. Equations~(\ref{eq:geongc}) and
     (\ref{eq:modetr}) shows $a_{\rm 
     geo} > r_{\rm tr}$ for $q > 1.4 \times 10^{-2}$ but less than the
     transitional  radius $r_{\rm m} \simeq 2.1 \times 10^4 r_{\rm G}
     q_{-1}^{5/7} \alpha_{-1}^{4/7} \dot{m}_{-3}^{-3/7}$ for $q\ga 2
     \times 10^{-3} \alpha_{-1}^{-20/11} 
     \dot{m}_{-3}^{15/11}$, where $\dot{m}_{-3} = \dot{m} / 10^{-3}$. 
     The secondary SMBH migrates inwards owing to the interaction with
     the standard accretion disk on a timescale given with
     equation~(\ref{eq:migtimred1})
\begin{equation}
  \tau_{\rm a} \simeq 3.88 \times 10^8 q_{-1}^{84/95}
  \alpha_{-1}^{-5/19} \delta_{-2}^{-10/19} .
\label{eq:ngctimev}
\end{equation}
    From equation~(\ref{eq:geocmbl}), the time derivative of geodetic
    precession because of the migration of the secondary interacting
    with a circumbinary accretion disk is
\begin{equation}
  {d P_{\rm geo} \over dt} \approx -0.21 q_{-1}^{-84/95}
  \alpha_{-1}^{5/19} \delta_{-2}^{10/19} . 
\label{eq:ngctmdrg}
\end{equation}
    The observed acceleration $\langle d P_{\rm pr} 
    /dt \rangle \simeq -0.19$ and equation~(\ref{eq:ngctmdrg}) suggest 
    a binary mass ratio 
    $q \approx 0.11 \alpha_{-1}^{25/84} \delta_{-2}^{25/42}$ and the
    secondary has mass $m \approx 3.8 \times 10^7 M_\odot
    \alpha_{-1}^{25/84} \delta_{-2}^{25/42}$. The
    binary separation is about $a \approx 7.0 \times 10^3 r_{\rm G} 
    \alpha_{-1}^{5/42} \delta_{-2}^{5/21} \simeq 0.23 \, {\rm
    pc}$ and $ a < a_{\rm h} \simeq 1.8 \times 10^4 r_{\rm G}
    \alpha_{-1}^{25/84} \delta_{-2}^{25/42}$. The results are
    insensitive to the disk parameters.  

    However, the secondary black hole should warp the standard thin
    disk and the warped disk would precess, probably leading to the
    precession of jet orientation with a timescale given with  
    equations~(\ref{eq:prec4}) and (\ref{eq:geongc}) 
\begin{equation}
  P_{\rm ap}  \simeq 1.08 \times 10^7 q_{-1}^{6/5} \alpha_{-1}^{5/3} 
  f_0^{-5/3} \delta_{-2}^{-10/3} \, {\rm yr} .
\label{eq:ngctimap}
\end{equation}
    If the precession timescale from the outer lobes to
    active jets is due to the precession of the warped disk,
    equation~(\ref{eq:ngctimap}) and the measured period $P_{\rm pr} = 
    2.72 \times 10^7 \, {\rm yr}$ give $q = 0.21 \alpha_{-1}^{-25/18}
    f_0^{25/18} \delta_{-2}^{25/9}$. The secondary has mass $m \approx
    7.2 \times 10^7 M_\odot \alpha_{-1}^{-25/18} f_0^{25/18}
    \delta_{-2}^{25/9}$ and the binary separation is about $a \approx
    9.1 \times 10^3 r_{\rm G} \alpha_{-1}^{-5/9} f_0^{5/9} \delta_{-2}^{10/9}
    \simeq 0.29 \, {\rm pc}$ and $a < a_{\rm h} \simeq 3.1 \times 10^4
    r_{\rm G} \alpha_{-1}^{-25/18} f_0^{25/18} \delta_{-2}^{25/9}$. 
    The secondary migrates inwards because of binary-disk interaction,
    leading to a time variation of jet precession timescale 
\begin{eqnarray}
  {d P_{\rm ap} \over dt} & \approx & - 1.4 \times 10^{-2}
  \alpha_{-1}^{1.88} f_0^{-1.88} \delta_{-2}^{-3.75} ,
  \label{eq:ngctimdrap} \\
	{d P_{\rm geo} \over dt} & \approx & - 0.11 
	\alpha_{-1}^{85/57} f_0^{-70/57} \delta_{-2}^{-110/57} .
\label{eq:ngctimdrgeo1}
\end{eqnarray}
    Equations~(\ref{eq:ngctimdrap}) and (\ref{eq:ngctimdrgeo1})
    implies that the time variations of jet precession cannot be
    detected because of the low observational accuracy. Therefore, the
    different precession timescale between the outer lobes and the
    active jets is most probably because of the different mechanism
    for jet precession. 

\subsection{Rapid acceleration of jet precession?}
\label{sec:acc}

    The argument for a steady precession in the object given by
    \citet{dunn06} is that if 
    the precession is speeding up, the acceleration would be very
    rapid and over the courses of about 1.5 rotation the precession
    timescale changes by around a factor of 10. However, our
    theoretical calculations suggest that a rapid
    acceleration is possible and there is no {\it a priori}
    requirement for constant precession timescale. In this section,
    we discuss the implications of a rapid acceleration of jet
    precession. We compute the precession timescale and the time
    derivatives for $n=0$ in Table~\ref{obser}. 
    From the averaged precession timescale from the
    Ancient to the outer lobes $\langle P_{\rm pr} \rangle \simeq
    15.23 \times 10^7 \, {\rm yr}$ and the averaged time derivative of
    the precession timescale from the Ancient bubbles to the outer
    lobes $\langle d P_{\rm pr} / dt \rangle \simeq -2.55$, we have
    the model-independent evolution timescale of SMBHB $\tau_{\rm ao}
    \simeq \left({\Lambda \over 1.5}\right) 8.96 \times 10^7 \, {\rm
    yr}$, which is about three times smaller than the timescale
    obtained with the assumption of a steady jet precession from the 
    Ancient bubbles through the Ghost bubble to the outer lobes. 
    While from the averaged precession timescale from the
    Ghost bubbles to the active jets $\langle P_{\rm pr} \rangle
    \simeq 5.27 \times 10^7 \, {\rm yr}$ and the averaged acceleration 
    of the precession from the Ghost bubbles to the active jets
    $\langle d P_{\rm pr} / dt \rangle \simeq -1.30$, we compute
    the model-independent evolution timescale of SMBHB $\tau_{\rm gj}
    \simeq \left({\Lambda \over 1.5}\right) 6.08 \times 10^7 \, {\rm
    yr}$.

    As the precession from the Ancient to the active jets is
    continuous without interruption when the activity of the object
    varies significantly, the possible mechanisms for jet 
    precession are the binary orbital motion and the geodetic
    precession of the primary. If the precession is due to binary
    orbital motion, the averaged precession timescale from the Ancient
    to the outer lobes $\langle P_{\rm pr} \rangle \simeq 15.23
    \times 10^7 \, {\rm yr}$ and equation~(\ref{eq:orb}) give an 
    averaged binary separation $a_{\rm ao} \simeq 5.01 \times 10^7
    r_{\rm G} \left({1 + q + {M_* \over M} \over 5 }\right)^{1/3}
    \simeq 1.46 \, {\rm Kpc}$. From the averaged time derivative of the 
    precession timescale from the Ancient bubbles through the
    Ghost bubbles to the outer lobes $\langle d P_{\rm pr} / dt
    \rangle \simeq -2.55$ and equation~(\ref{eq:acorbdy}), we obtain
    the binary mass ratio $q_{\rm ao} \approx 0.92 \left({1 + q + {M_*
    \over M} \over 5}\right)^{4.02/9}$. Meanwhile, the
    averaged precession timescale from the Ghost bubbles to the active
    jets $\langle P_{\rm pr} \rangle \simeq 5.27 \times 10^7 \, {\rm
    yr}$ and equation~(\ref{eq:orb}) give an averaged binary 
    separation $a_{\rm gj} \simeq 2.47 \times 10^7 r_{\rm G} \left({1
    + q + {M_* \over M} \over 5 }\right)^{1/3} \simeq 0.80 
    \, {\rm Kpc}$. From the averaged time derivative of the 
    precession timescale from the Ghost bubbles through the
    outer lobes to the active jets $\langle d P_{\rm pr} / dt
    \rangle \simeq -1.30$ and equation~(\ref{eq:acorbdy}), we have
    binary mass ratio $q_{\rm gj} \approx 0.60 \left({1 + q + {M_*
    \over M} \over 5}\right)^{4.02/9}$. The two averaged time
    derivatives give a consistent mass ratio and suggest a major
    merger with $\langle q \rangle \approx 0.76 \left({1 + q + {M_*
    \over M} \over 5}\right)^{4.02/9}$. From galactic dynamics
    \citep{binney87}, it is expected that the evolution timescale of a
    SMBHB due to dynamic friction is nearly proportional to the
    separation, $\tau_{\rm a} \propto a$. Our results give 
    $\tau_{\rm ao} / \tau_{\rm gj} \simeq 1.5$ and $a_{\rm
    ao} / a_{\rm gj} \simeq 2.0$, which are consistent with the
    predications very well and give an averaged dynamic friction 
    velocity 
\begin{equation}
  \langle { d a \over dt} \rangle \simeq - 1.54 \times 10^6 \left({1
    + q + {M_* \over M} \over 5 }\right)^{1/3} \, {\rm cm/s} 
\end{equation}
    where $M_*$ is the stellar mass inside the binary orbit at 
    separation $a\sim 1 \, {\rm Kpc}$. 

    The alternative for jet precession independent of source
    activity and the accretion is the binary geodetic precession. 
    From equation~(\ref{eq:precbbr}), to obtain the averaged precession
    timescale from the Ancient to the outer lobes, we have the
    binary separation 
    $a_{\rm geo} \simeq 1.24 \times 10^4 r_{\rm G} q_{-1}^{2/5}$. 
    Because $ a_{\rm geo} \la a_{\rm h} \simeq 1.77 \times 10^4 r_{\rm
    G} q_{-1}/ (1 +q)$, the binary is hard and the secondary may
    migrate inward when it interacts with a light standard disk,
    leading to the acceleration of jet precession
\begin{equation}
  {d P_{\rm geo} \over dt} \simeq -0.410 \dot{m}_{-3}^{14/19}
  q_{-1}^{-84/95} \alpha_{-1}^{5/19} \delta_{-2}^{10/19} ,
\label{eq:ngcrptmge}
\end{equation}
    where we have used accretion rate $\dot{m}\sim 10^{-3}$ and
    equation~(\ref{eq:geocmbl}). Equation~(\ref{eq:ngcrptmge}) and the
    measured time derivative give $q \simeq 
    1.3 \times 10^{-2} \dot{m}_{-3}^{5/6} \alpha_{-1}^{25/84}
    \delta_{-2}^{25/42}$. Similarly, from the observation of jet
    precession from the Ghost bubbles to active jets, we have $a_{\rm
    geo} \simeq 8.13 \times 10^3 r_{\rm G} q_{-1}^{2/5}$ and $q \simeq 
    0.97 \times 10^{-2} \dot{m}_{-3}^{5/6} \alpha_{-1}^{25/84}
    \delta_{-2}^{25/42}$. Although the estimated  mass ratios are
    consistent with each other within the uncertainties of accretion
    rate, the hardening of a hard SMBHB depends on the accretion and
    the migration of the secondary should stop when the source becomes
    dormant and the accretion disk becomes ADAF. Even if the 
    migration could happen when the source is luminous and forms the
    bubbles, the time scale to form a hard binary with mass ratio
    $q\sim 10^{-2}$ is $\tau \ga 2\times 10^9 \, {\rm yr}$ and is
    inconsistent with the scenario of recent merger. Therefore, the
    scenario of geodetic precession for jet precession is less
    favorable.

\section{Discussions and conclusions}
\label{sec:dis}

    SMBHBs are expected by the hierarchical galaxy formation model and
    may have been observed in many AGNs. Jet precession observed in
    many AGNs is one of the observational evidences. In this paper, we 
    start our work with the discussion of different mechanisms for the
    jet precession, 
    including (1) the geodetic precession of spin axis of central
    primary SMBH around total angular momentum, (2) the orbital motion
    of the SMBH ejecting plasma jets, (3) the inner disk precession
    because of the tidal interaction of an inclined secondary black
    hole, (4) the precession of a circumbinary  disk warped by the
    SMBHB, and (5) the disk precession because of Bardeen-Peterson
    effect. The precession of a circumbinary  disk warped by a SMBHB is
    discussed first time. We did not discuss the precession model due
    to disc instability \citep{pringle97}, because it suggests a
    stochastic precession rather than a regular precession and is 
    inconsistent with the observations of jet precession in most AGNs.
    Although Bardeen-Peterson effect does not directly connect to the
    presence of SMBHB, the origin of misalignment between the
    rotating central black hole and the accretion disk may be due to
    the interaction of accretion disk and an inclined SMBHB. When the
    inner disk region becomes misaligned with the binary orbital plane
    owing to the Bardeen-Peterson effect, the tidal interaction of the
    secondary to the warped inner disk also leads to jet
    precession. 

    In these scenarios for jet precession, the precession
    timescale ranges from order of years to much longer than $10^8\,
    {\rm yr}$, depending on the parameters of SMBHB and accretion
    disk. However, the parameters are very difficult to determine and
    the observations of jet precession timescale cannot give restrict
    constraints on the models and the parameters, as they are
    degenerate. Therefore, we suggested to observe one more quantity,
    the time variations of jet precession timescale, to resolve the
    parameters. We calculated the time variation of jet
    precession in different models and showed that jet
    precession is always accelerated in an evolving SMBHB
    system. The acceleration of jet precession is related to the
    evolution timescale of SMBHB with $ {dP_{\rm pr} \over dt}
    \simeq - \Lambda {P_{\rm pr} \over \tau_{\rm a}}$, resulting from  
    the fact that all SMBHB models for jet 
    precession predicate a relation $P_{\rm pr} \propto
    a^\Lambda$ with $\Lambda > 0$ and that a SMBHB in galactic nuclei
    never gets softer. The parameter $\Lambda$ slightly depends on
    model with $4/3 \leq \Lambda \leq 3$. Our investigations also show
    that jet precession because of Bardeen-Peterson effect is
    decelerated with AGN evolution. Our results suggest that the sign
    of the time derivative of precession timescale can be used to
    identify SMBHB models from the others. 

    Our calculations show that the time variation of jet precession
    is proportional to the timescale ratio of jet precession
    and SMBHB evolution. We analytically estimated the evolution
    timescale of SMBHBs at different evolution stage and the time
    variation for jet precession in different models, based on our
    current knowledge of SMBHBs. Our calculations show that for an
    un-bound SMBHB the mechanism for jet precession is the orbital
    motion and the quick binary evolution because of galactic
    dynamic friction leads to around 20 \% or higher acceleration rate
    of jet precession timescale. For a bound SMBHB system, jet
    precession could be due to geodetic precession of the rotating
    primary black hole, disk precession because of tidal interaction
    between a standard accretion disk and the secondary, and the
    binary orbital motion. At this stage, the evolution timescale of
    SMBHB depends on the inner surface brightness profiles of
    galaxies and is estimated with an asymptotic analytic relation of
    the binary hardening timescale and the separation given by
    \citet{yu02}. Although the 
    estimate is very rough, our results suggest that the evolution
    timescale of SMBHB is several order of magnitude shorter than the
    geodetic precession timescale and longer than the binary orbital
    period. So, the geodetic precession is not significant and the
    time variation of orbital motion is difficult to measure. However,
    if the jet precession is due to the tidal interaction of the
    secondary black hole and an inner misaligned accretion disk, the 
    acceleration rate of jet precession could be a order of 10 \% or
    higher. 

    When a SMBHB becomes hard and stalls, jet precession may be steady
    for a timescale longer than the Hubble time. Because the migration
    of the secondary SMBH due to the interaction with an ADAF is
    negligible \citep{narayan00,liu04}, a nearly steady
    precession jet may be possible if a SMBHB interacts with an ADAF
    or the precession is due to Bardeen-Peterson effect. The
    fundamental difference between the two scenarios is that the
    accretion disk is a geometrically thin standard disk in the later
    but geometrically thick ADAF in the former. The accretion mode
    depends on the relative accretion rate $\dot{m}$, which one could
    infer by estimating the bolometric luminosity and central black
    hole mass. However, it is most probable that a SMBHB interacts
    with a standard disk either massive or light. We compute the time
    variation of jet precession because of SMBHB-accretion disk
    interaction and show that in both cases the binary-disk
    interaction would lead to a significant acceleration of jet
    precession: the acceleration is significant for jet precession 
    because of tidal interaction of the secondary and a massive disk
    but both of geodetic precession and warped circumbinary disk
    precession in the case of non-massive disk, depending on the
    parameters of the binary system and the accretion disk.

    When the evolution of a SMBHB is dominated by the gravitational
    wave radiation, the binary separation is about hundreds of
    Schwarzschild radius or less. If a jet ejects from the central
    black hole, it precesses because of the black hole geodetic
    precession, of the tidal interaction of binary and inner
    misaligned disk, of binary orbital motion, and of the precession
    of warped circumbinary disk. Our calculations show that the
    precession of a warped circumbinary disk is strongly 
    accelerated owing to the migration of the secondary black hole
    because of gravitational wave radiation. The acceleration of jet
    precession because of geodetic precession is also very significant
    for a SMBHB with non-zero eccentricity. 

    When we calculate the jet precession timescale and its time
    variation, we have assumed that disk precession is rigid-like and
    the jet precession is directly related to it. Although almost all
    the disk precession models for jet precession in the literature
    adopted the same assumption and have successfully explained the
    jet precession in some AGNs and micro-quasars (e.g. SS433), this
    assumption need more discussions. Whether the assumption of a
    rigid body like precession is valid or not depends on the warp
    transfer in the disk. As we have discussed in
    Section~\ref{sec:nmassdisk}, the transportation of warps in disk
    depends on the vertical shear viscosity and the transfer timescale 
    at the transition radius between the warped and unperturbed disk
    regions is on the same order of the precession timescale both for
    the Bardeen-Peterson effect and the warped circumbinary light disk
    \citep[e.g.][]{natarajan98,liu04}. Therefore, the rigid body like
    approximation for disk precession is correct on the zero order of
    magnitude. As the jet precession and its time derivative depend on
    disk characters in a similar way, the relationship of the ratio
    of the precession timescale and its variation rate, $P_{\rm
    pr} / \dot{P}_{\rm pr} = - \left[P_{\rm pr} / \left({\partial
    P_{\rm pr} \over \partial a}\right)\right] (\tau_{\rm a} /a)$, and
    the SMBHB evolution timescale $\tau_{\rm a}$ would be expected to be
    insensitive to how warps transfer in the disk.

    Following our theoretical investigation on the acceleration of jet
    precession, we discussed the implications of the differential
    observations of jet precession in NGC1275 (3C84), a recent-merger 
    radio galaxy. The differential jet precession have been measured
    between four different components in order of formation: ancient
    bubbles, ghost bubbles, outer lobes, and the active jets. Between
    the formation of different components, the activity of the object
    becomes very weak or the source is dormant. The precession
    timescale are significantly decreased with time among the
    different components. \citet{dunn06} assumed a steady jet
    precession and the acceleration of jet precession just because of
    the missing of several cycles between adjacent
    components. However, even under this assumption the acceleration
    of jet precession from the ghost bubbles to the active jets is
    still significant. Because the precession is steady when the
    source activity changes dramatically, the mechanism for the
    precession is independent of the accretion and thus most likely of
    the geodetic precession or orbital motion of SMBHB. Under the
    assumption of steady jet precession, we discussed the two possible
    mechanisms. Our discussions suggest that if the precession is due
    to the binary orbital motion, the SMBHB should have a too small
    mass ratio ($q \la 2\times 10^{-2}$) and the acceleration of jet 
    precession from the ghost bubbles through outer lobes to the
    active jets cannot be explained reasonably. Our results show that
    the steady precession from the ancient bubbles to the outer lobes
    is probably due to the geodetic precession and the jet precession
    from the outer lobes to active jets may be due to the precession
    of a warped circumbinary light standard thin disk. In this
    scenario, SMBHB formed in a major merger with mass ratio $q = 0.21
    \alpha_{-1}^{-25/18} f_0^{25/18} \delta_{-2}^{25/9}$ and the
    binary has a separation $a \approx 9.1 \times 10^3 r_{\rm G}
    \alpha_{-1}^{-5/9} f_0^{5/9} \delta_{-2}^{10/9} \simeq 0.29 \,
    {\rm pc}$. The predicated acceleration of jet precession is about
    a few to ten percent and may have not yet been observed because of
    the low observational accuracy. 

    Like what our theoretical investigations show that there is no
    {\it a priori} requirement for a steady jet precession, we
    discussed the implications that if a continuous rapid acceleration
    of precession from the ancient bubbles to the active jets has
    indeed been observed. Our results show that in this case the
    mechanism for jet precession is the orbital motion and that the
    rapid acceleration of jet precession is due to the rapid evolution
    of SMBHB because of galactic dynamic friction. The calculations
    give a galactic dynamic friction evolution timescale $\tau_{\rm a}
    \approx (6-9)\times 10^7 \, {\rm yr}$ and an averaged dynamic
    friction velocity ${ d a \over dt} \approx - 1.54 \times 10^6 \,
    {\rm cm/s}$. The SMBHB forms in the major galaxy merger with an
    averaged black hole mass ratio $q \approx 0.76$ and has a
    separation $a \approx 0.8 - 1.46 \, {\rm Kpc}$. 

    As our conclusions, we discussed the scenarios for jet
    precession in AGNs and calculated the time derivatives of the
    precession timescale. Our calculations show that jet precession
    is accelerated in SMBHB models but nearly steady in the
    Bardeen-Peterson effect scenario. We analytically computed the
    predicated acceleration of jet precession in the evolution of
    a SMBHB from unbound to gravitational-wave-dominated stages and
    showed that the time variation is significant and can be detected
    easily. One can estimate the evolution timescale and mass ratio of
    SMBHB, and the parameters of accretion disk in AGNs by measuring
    the central black hole mass, the accretion rate, the jet
    precession timescale, and its time derivative. If we have
    observations of jet precession acceleration of a sample of
    radio sources, we can test the hierarchical galaxy formation model
    and the galactic dynamics. We can also estimate the fraction of
    SMBHB that can get coalesced quickly and give rise to
    gravitational wave radiation bursts.

\acknowledgments

    We are grateful to D.N.C. Lin, J. Magorrian, J.F. Lu, and X.-B. Wu
    for helpful discussions and comments. Many thanks are due to the
    referee for constructive comments, which made us to improve
    the presentation of the paper significantly. This work is
    supported by the National Natural Science Foundation of China
    (No. 10573001).

\end{document}